# Suppression of Spin Pumping at Metal Interfaces

Youngmin Lim[1,a)], Bhuwan Nepal[2], David A. Smith[1,b)], Shuang Wu[1,c)], Abhishek Srivastava[2], Prabandha Nakarmi[2], Claudia Mewes[2], Zijian Jiang[1], Adbhut Gupta[1], Dwight D. Viehland[3], Christoph Klewe[4], Padraic Shafer[4], In Jun Park[5], Timothy Mabe[5], Vivek P. Amin[5], Jean J. Heremans[1], Tim Mewes[2], Satoru Emori[1,*]

1. Department of Physics, Virginia Tech, Blacksburg, Virginia 24061, USA

2. Department of Physics and Astronomy, The University of Alabama, Tuscaloosa, Alabama 35487, USA

3. Department of Materials Science and Engineering, Virginia Tech, Blacksburg, Virginia 24061, USA

4. Advanced Light Source, Lawrence Berkeley National Laboratory, Berkeley, California 94720, USA

5. Department of Physics, Indiana University - Purdue University Indianapolis, Indianapolis, Indiana 46202, USA

a) Present address: Micron Technology, Boise, Idaho 83716, USA.

b) Present address: HRL Laboratories, Malibu, California 90265, USA.

c) Present address: Western Digital Corporation, San Jose, California 95119, USA.

*Author to whom correspondence should be addressed: semori@vt.edu

**Abstract**

An electrically conductive metal typically transmits or absorbs a spin current. Here, we report on evidence that interfacing two metal thin films can suppress spin transmission and absorption. We examine spin pumping in ferromagnet/spacer/ferromagnet heterostructures, in which the spacer – consisting of metallic Cu and Cr thin films – separates the ferromagnetic spin-source and spin-sink layers. The Cu/Cr spacer largely suppresses spin pumping – i.e., neither transmitting nor absorbing a significant amount of spin current – even though Cu or Cr alone transmits a sizable spin current. The antiferromagnetism of Cr is not essential for the suppression of spin pumping, as we observe similar suppression with Cu/V spacers where V is a nonmagnetic analogue of Cr. We speculate that diverse combinations of spin-transparent metals may form interfaces that suppress spin pumping, although the underlying mechanism remains unclear. Our work may stimulate a new perspective on understanding and engineering spin transport in metallic multilayers.

# I. Introduction

The flow of spin angular momentum, i.e., spin current, plays key roles in spintronic phenomena. In particular, *pure* spin currents – which are not accompanied by net charge flow – may enable novel devices that surpass the limitations of spin-polarized charge currents [1,2]. It is especially crucial to understand the fundamentals of pure spin currents in metallic multilayers (heterostructures) comprising practical spintronic devices [2,3].

Spin pumping is an oft-used method to study pure spin currents [4,5] – for instance, in spin-valve-like heterostructures consisting of a ferromagnetic spin *source*, a *spacer*, and a ferromagnetic spin *sink* [Fig. 1]. In this method, microwave-driven ferromagnetic resonance (FMR) excites the magnetization in the spin source, which pumps an ac pure spin current that propagates into the adjacent layer. Prior spin pumping experiments have often been performed on heterostructures with Cu as the spacer [6–9], as illustrated in Fig. 1(a). In this case, the spin current is transmitted through the spacer with practically no decay, due to the long spin diffusion length of $\gg$100 nm in Cu [10,11]. The transmitted spin current is then absorbed in the spin sink, leading to a nonlocal loss of nonequilibrium spin angular momentum from the spin source. This loss manifests in *spin-pumping damping* [4,5], an enhanced damping $\Delta\alpha$ over the intrinsic Gilbert damping parameter $\alpha_0$ of the ferromagnetic source.

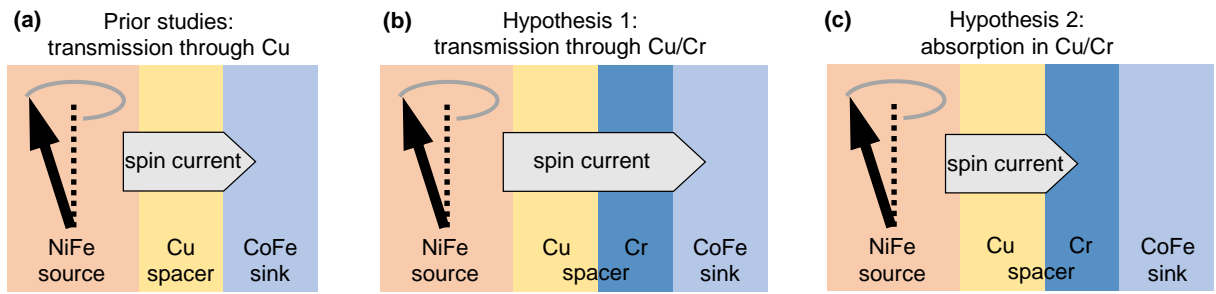

FIG 1. Simple schematics of spin-valve-like heterostructures, in which FMR in the NiFe source pumps a pure spin current. (a) Transmission of the pumped spin current through the Cu spacer, which is well-established from such prior studies as Refs. [6–9]. The spin current is absorbed quickly in the ferromagnetic CoFe sink. (b,c) Two hypothesized scenarios for spin transport in heterostructures incorporating an additional Cr layer in the spacer: the spin current may be (b) transmitted through the bilayer Cu/Cr spacer or (c) absorbed in the Cu/Cr spacer (or Cr layer). Neither of these hypotheses turns out to match our experimental results.

Our present study aims to reveal how spin pumping is affected by incorporating a thin layer of another elemental metal – such as Cr – in the spacer of a heterostructure. Cr is an interesting choice, in part because it is a well-known elemental antiferromagnet with a rich assortment of magnetic order [12,13]. From this viewpoint, our study was originally intended to contribute to the growing discipline of antiferromagnetic spintronics, which had investigated spin transport in antiferromagnetic alloys and compounds [14–19]. Studying Cr-based heterostructures is also timely for spin-orbitronics [2,20], as several groups have reported significant spin and orbital Hall effects in Cr [21–27].

More crucially, spin transport in Cr is intriguing because contradictory findings have been reported. On one hand, an experimental study reports a spin diffusion length of $\approx$13 nm in Cr [21], which – though much shorter than in Cu – is several times greater than in other transition metals (e.g., W, Ta, Pt) [28–30] and metallic antiferromagnets (e.g., IrMn, FeMn) [14–18]. Considering Cr's low electrical resistivity (bulk room-temperature value $\approx$13 μΩ cm) and low atomic number ($Z = 24$, hence presumably weak spin-orbit coupling to decohere spins), it appears reasonable that spin currents can be transmitted over a $\gtrsim$10-nm length scale in Cr. On the other hand, a separate study reports a much shorter spin diffusion length of ≈ 2 nm in Cr [22]. In this case, even ultrathin Cr should efficiently absorb a spin current. Thus, how an additional thin Cr layer affects spin transport in magnetic heterostructures [Fig. 1(b,c)] remains an open question. Moreover, spin transport in Cr could be anisotropic – e.g., dependent on the propagating spin polarization with respect to a certain crystallographic axis [31]. It is then instructive to examine how the crystalline structure of Cr influences spin pumping.

Here, we investigate pure-spin-current transport in magnetic multilayers incorporating thin-film Cr of thickness ≲ 10 nm. We primarily study spin pumping in spin-valve-like heterostructures, illustrated in Fig. 1, each consisting of a NiFe spin *source*, a Cu/Cr *spacer*, and a CoFe spin *sink*. We initially hypothesized two scenarios:

> *Hypothesis 1* [Fig. 1(b)]: The spin current is transmitted through the Cu/Cr spacer and is absorbed in the CoFe sink. The spin absorption results in spin-pumping damping.
>
> *Hypothesis 2* [Fig. 1(c)]: The spin current is absorbed in the Cu/Cr spacer. The spin absorption in this case also results in spin-pumping damping, even without the CoFe sink – because Cu/Cr effectively behaves as a sink in this case.

As it turns out, our experimental observations do not match either of these hypothesized scenarios. In fact, inserting even an ultrathin (~1 nm) layer of Cr *suppresses* spin pumping – i.e., most of the spin current is *neither transmitted nor absorbed* in the Cu/Cr spacer. This finding is rather surprising, especially as we verify that Cr alone (not interfaced with Cu) transmits the spin current. Thus, we deduce that the suppression of spin pumping emerges from the Cu/Cr interface. We also find that the suppression of spin pumping does not require antiferromagnetic order in Cr; similar suppression is observed with Cu/V spacers without any antiferromagnetism. Hence, this peculiar effect of suppressed spin pumping may arise at the interfaces of other nonmagnetic metals. Our findings have the potential to cultivate a new fundamental perspective on spin transport across metal interfaces.

## II. FILM GROWTH AND STRUCTURE

### A. Rationale for the Heterostructures

To examine the influence of crystalline structure on spin transport, we have grown two series of NiFe/Cu/Cr/(Co)Fe heterostructures:

(1) those incorporating *epitaxial* Cr, verified in Sec. II-B to be (001)-oriented, grown on top of epitaxial (Co)Fe on (001)-oriented single-crystal $MgAl_2O_4$ (MAO) [Fig. 2(a)], and

(2) those incorporating *polycrystalline* Cr, verified in Sec. II-B to be (110)-textured, grown on top of other polycrystalline film layers on Si substrates with $SiO_2$ native oxide [Fig. 2(b)].

These samples were grown by dc magnetron sputtering with a base pressure of ≲ 5×10$^{-8}$ Torr and an Ar sputtering gas pressure of 3 mTorr. In all heterostructures, the composition of the NiFe spin source is $Ni_{80}Fe_{20}$ (permalloy). The (Co)Fe spin sink is $Co_{25}Fe_{75}$ in most cases, but we also use elemental Fe for a few samples. The factor of ≈2 greater saturation magnetization for (Co)Fe compared to NiFe results in a large separation between the FMR conditions of the two ferromagnets. As such, we can readily extract the FMR linewidth of the NiFe spin source that is well distinguished from the FMR spectrum of the (Co)Fe spin sink.

Figure 2(a) depicts the heterostructure incorporating epitaxial Cr interfaced with epitaxial (Co)Fe. The MAO substrate is well lattice-matched to BCC-(Co)Fe to within ≈0.4% [32]. After pre-annealing the MAO substrate at 600 °C for 2 hours to drive off surface contaminants, the 4-nm-thick (Co)Fe layer was deposited at a substrate temperature of 200 °C. The Cr layer of thickness 0-12 nm was grown on top of (Co)Fe at 150 °C; the somewhat lower substrate temperature was intended to decrease intermixing between the (Co)Fe and Cr layers. Then, the substrate was cooled to room temperature; during this cooling process, the background pressure in the deposition chamber was ≲ 5×10$^{-8}$ Torr, preventing oxidation of the Cr surface. Finally, the Cu (5 nm), NiFe (10 nm), and Ti (3 nm) layers were deposited. The Ti capping layer protects the underlying stack from oxidation when the sample was taken out of the deposition chamber for measurements at ambient conditions. We remark that having the NiFe spin source at the bottom would have been preferable to minimize extrinsic FMR linewidth broadening [33,34], e.g., caused by film roughness propagated from the underlying layers. Yet, in this samples series [Fig. 2(a)], the NiFe spin source must necessarily be on top to allow for epitaxial growth of the (Co)Fe and Cr layers. We find negligible extrinsic FMR linewidth broadening in the NiFe spin source so long as NiFe is grown on Cu on top of the epitaxial (Co)Fe/Cr underlayers, thereby permitting reliable characterization of spin pumping.

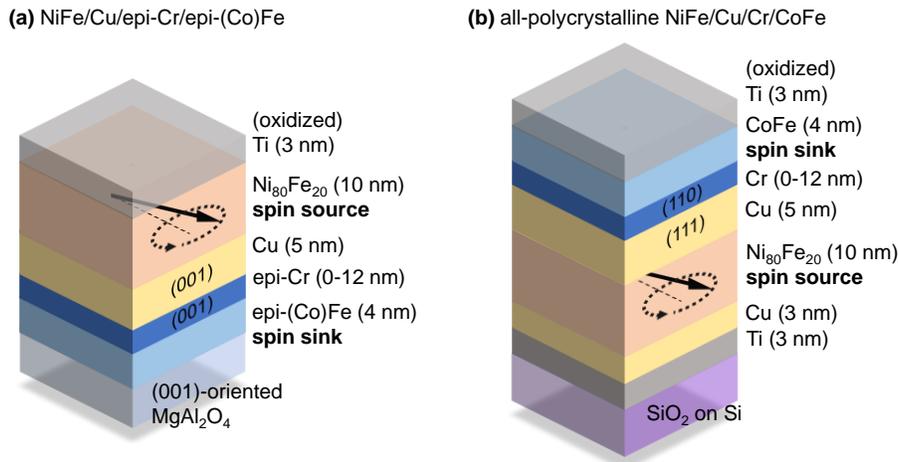

FIG 2. Schematics of heterostructures primarily investigated in this work (a) based on epitaxial Cr and (Co)Fe grown on (001)-oriented single-crystal MAO and (b) comprised entirely of polycrystalline layers grown on SiO$_2$ on Si. The out-of-plane crystallographic orientations of the Cu/Cr spacers are indicated.

Figure 2(b) depicts the heterostructure in which all constituent layers are polycrystalline. These all-polycrystalline stacks were grown with the Si-SiO$_2$ substrate at room temperature. Since this sample series [Fig. 2(b)] does not involve the epitaxial growth of Cr, the NiFe spin source was grown on the bottom side of the heterostructure to reduce the possible influence from underlayer roughness. The NiFe layer was seeded by Ti(3 nm)/Cu(3 nm) to minimize extrinsic FMR linewidth broadening [35]. As in the epitaxial series, each film stack in the polycrystalline series was capped with 3-nm-thick Ti for protection against oxidation.

In both sample series illustrated in Fig. 2, the NiFe source and Cr are separated by a 5-nm-thick spacer of diamagnetic Cu. The Cu spacer eliminates potential complications that might arise from directly interfacing Cr with NiFe, such as proximity-induced magnetism [36–38] or magnon coupling between NiFe and antiferromagnetic Cr [39–41]. NiFe grown directly on top of epitaxial Cr shows indication of anisotropic two-magnon scattering [42,43], which complicates quantification of spin-pumping damping. By contrast, two-magnon scattering is largely absent in NiFe seeded by polycrystalline Cu.

In principle, (Co)Fe could be used as the spin source and NiFe as the spin sink. However, spin pumping measurements become complicated with a (Co)Fe source, due to pronounced non-Gilbert contributions to the FMR linewidth [33,34]. In 4-nm-thick (Co)Fe, we observe a large zero-frequency linewidth (e.g., ≳1 mT), sometimes accompanied by a nonlinear frequency dependence of the linewidth, varying from sample to sample. Such complicated behavior may arise from two-magnon scattering from magnetic inhomogeneity [44,45], perhaps underpinned by non-uniform strain or interfacial roughness. We were thus unable to quantify the Gilbert damping parameter for the thin (Co)Fe layers reliably. In contrast, we find negligible zero-frequency linewidths of only ~0.1 mT and a linear trend of linewidth vs frequency for NiFe layers (especially those grown on top of Cu). That is, the FMR linewidths of such NiFe layers are less vulnerable to the spurious two-magnon scattering contribution, likely because the small magnetostriction of $Ni_{80}Fe_{20}$ reduces magnetic inhomogeneity. In this study, we exclusively focus on NiFe as the spin source, as it permits straightforward quantification of Gilbert damping that is essential for probing spin pumping.

## B. Crystallographic Orientations of the Heterostructures

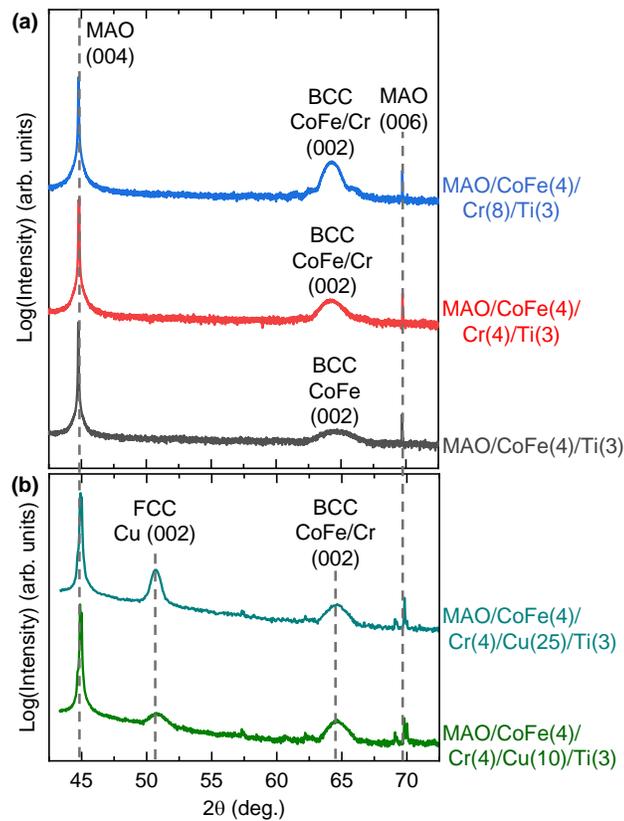

FIG 3. XRD spectra for (a) samples with 0-, 4-, and 8-nm-thick Cr grown on top of epitaxial CoFe and (b) samples with 4- and 10-nm-thick Cu grown on top of epitaxial CoFe/Cr. In both (a) and (b), the (001)-oriented MAO substrate allows for epitaxial growth of CoFe, and the 3-nm-thick Ti capping layer protects the underlying films from oxidation. Also note that (a) was acquired with a Panalytical high-resolution diffractometer, whereas (b) was acquired with a Bruker powder diffractometer, hence resulting in different backgrounds in the XRD spectra.

We have compared the crystallographic orientations of Cr in the epitaxial and polycrystalline series through 2θ-ω x-ray diffraction (XRD) measurements. Figure 3 summarizes our XRD results for epitaxial Cr, along with the Cu layer interfaced with it. We confirm that 4-nm-thick BCC CoFe is (001)-oriented, as evidenced by the (002) film diffraction peak [Fig. 3(a)]. With the addition of Cr on top of CoFe, the (002) film peak becomes taller, indicating that the BCC Cr layer is also (001)-oriented. This is unsurprising considering the similar bulk lattice parameters of BCC $Co_{25}Fe_{75}$ (≈0.287 nm) and BCC Cr (≈0.291 nm). In Fig. 3(b), we show XRD spectra for samples with Cu deposited at room temperature (hence presumed to be polycrystalline) on top of epitaxial CoFe/Cr bilayers. A diffraction peak corresponding to the (002) plane of FCC Cu is evident. Thus, the Cu layer develops a (001) orientation on top of (001)-oriented epitaxial Cr, despite the large difference in lattice parameter between FCC Cu (≈0.361 nm) and BCC Cr.

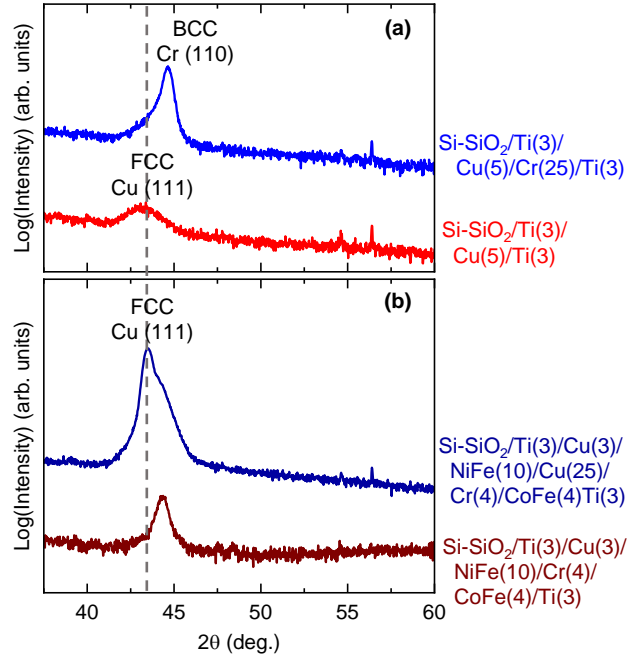

FIG 4. XRD spectra for all-polycrystalline samples. (a) Comparison of the crystallographic texture for Cu and Cr. (b) Verification of the (111) texture of Cu grown on top of NiFe. Note that these all-polycrystalline samples are seeded by Ti/Cu on Si substrates with native $SiO_2$, and capped by Ti. The large Cr and Cu thicknesses of 25 nm in (a) and (b), respectively, facilitates disentangling the Cr and Cu diffraction peaks from the rest of the film stack.

Figure 4 shows XRD results that reveal the structures of Cu and Cr in our polycrystalline samples. In Fig. 4(a), we see that the polycrystalline Cr layer has a (110) texture when deposited on top of (111)-textured Cu. Figure 4(b) further confirms that a Cu layer grown on a Ti/Cu/NiFe stack maintains a (111) texture. The polycrystalline film layers grown on amorphous $SiO_2$ (without any templating from a single-crystal substrate) favor closest-packed planes: (111) for FCC Cu and (110) for BCC Cr.

Some XRD spectra in Figs. 3(b) and 4 show a small peak at $2\theta \approx 57°$. Diffraction peaks near that range of $2\theta$ have been reported for $Cr_2O_3$ [46]. However, a peak at $2\theta \approx 57°$ is still present even in Si-$SiO_2$/Ti/Cu/Ti (Fig. 4(a)) without any Cr. Moreover, such a peak is absent for MAO/CoFe/Cr/Ti samples measured with a different diffractometer (see caption for Fig. 3). We attribute the peak at $2\theta \approx 57°$ to an instrumental background, rather than oxidized Cr.

To summarize the above XRD results, we find different crystallographic orientations of Cu/Cr for the epitaxial series [Fig. 3] and the polycrystalline series [Fig. 4]. Namely, the spacer in the epitaxial series consists of Cu(001)/Cr(001) [Fig. 3], whereas that in the polycrystalline series consists of Cu(111)/Cr(110) [Fig. 4]. The epitaxial and polycrystalline series hence provide distinct model systems to examine the role of Cu/Cr structure in spin transport. Nevertheless, as shown in the following section, we find that the structurally different Cu/Cr spacers both yield significant suppression of spin pumping.

## III. RESULTS AND DISCUSSION

### A. Measurement of Spin-Pumping Damping

We employ broadband FMR spectroscopy to study spin transport in our heterostructures by monitoring nonlocal damping enhancement of the spin source [4,5]. In the following discussion of spin pumping, we represent each heterostructure with the notation "NiFe/spacer/sink," such that the spin current propagates from "left" (NiFe source) to "right" (sink). Unless otherwise specified, our notation omits the substrate and the seed and capping layers for simplicity; Section II-A (in particular, Fig. 2) describes the constituent layers of the heterostructures.

Our spin pumping measurements are performed at room temperature, except for those in Sec. III-E that extend to 10 K. The sample is placed film-side down on a coplanar waveguide to excite resonant magnetic precession in the NiFe spin source. A magnetic field from an electromagnet is applied along the film plane. The magnetic precession in the NiFe spin source pumps an ac pure spin current into the adjacent layers.

Any spin current transmitted through the spacer is absorbed by the ferromagnetic (Co)Fe spin sink [7,47]. The spin absorption in the (Co)Fe sink constitutes a loss of spin angular momentum emitted by the NiFe source, hence increasing Gilbert damping in the NiFe layer [4,5]. Alternatively, some of the spin currents could be absorbed within the Cu/Cr spacer, which would also enhance damping in the NiFe source. Therefore, the additional damping $\Delta\alpha$ from spin absorption (outside of the NiFe source) is

$$\Delta\alpha = \alpha - \alpha_0, \quad (1)$$

i.e., the difference between the total measured Gilbert damping parameter $\alpha$ and the baseline intrinsic Gilbert damping parameter $\alpha_0$ of NiFe.

From field-swept FMR measurements performed at frequencies $f$ = 2-22 GHz (additional details available in Refs. [18,48]), we extract $\alpha$ by linearly fitting the $f$ dependence of the half-width-at-half-maximum FMR linewidth $\Delta H$ via

$$\mu_0 \Delta H = \mu_0 \Delta H_0 + \frac{2\pi}{\gamma}\alpha f, \quad (2)$$

where $\Delta H_0$ is the zero-frequency linewidth of ≲0.1 mT attributed to small inhomogeneous broadening and $\gamma/(2\pi) = 29.5$ GHz/T is the gyromagnetic ratio typical for $Ni_{80}Fe_{20}$.

Figure 5 shows representative results for the frequency dependence of the FMR linewidth. NiFe without a spin sink show $\alpha = \alpha_0 \approx 0.007$ [Fig. 5(a)], in good agreement with previously reported room-temperature damping parameters of $Ni_{80}Fe_{20}$ [48,49]. In the following, we use $\alpha_0 = 0.00710 \pm 0.00015$ obtained by averaging results on films from different deposition runs. The stack structure of these baseline samples is Si-SiO$_2$ (substrate)/Ti/Cu/NiFe/Cu/Ti. We note that Ti and Cu contribute negligibly to $\Delta\alpha$. The spin current is unable to enter 3-nm-thick Ti that is likely oxidized (leading to high resistivity ~1000 μΩ cm) by being directly interfaced with the oxide substrate or ambient air. The spin diffusion length in Cu [10,11] is much greater than the Cu spacer thickness here, such that spin backflow in the Cu layer cancels the spin current pumped out of the NiFe source [4,10,11]. Additional baseline samples of NiFe on epitaxial underlayers (i.e.,

MAO (substrate)/epi-Cr/Cu/NiFe/Ti) show two-magnon scattering, but the baseline Gilbert damping parameter of these samples is also deduced to be $\alpha_0 \approx 0.0071$ [see Supplementary Material].

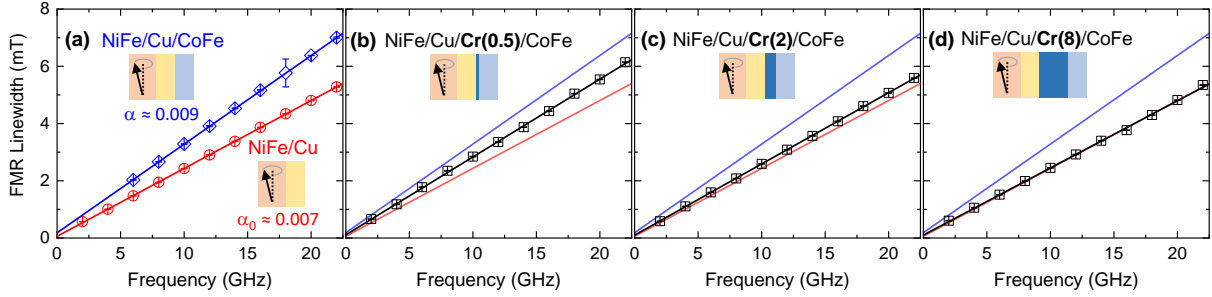

FIG 5. Frequency dependence of the half-width-at-half-maximum FMR linewidth for (a) NiFe/Cu/CoFe (with CoFe as the spin sink) and NiFe/Cu (without a spin sink), as well as NiFe/Cu/Cr/CoFe with Cr insertion layer thicknesses of (b) 0.5 nm, (c) 2 nm, and (d) 8 nm. The ferromagnet/spacer/ferromagnet heterostructures shown here are based on epitaxial CoFe grown on MAO substrates (i.e., the heterostructure illustrated in Fig. 2(a)).

The NiFe/Cu/CoFe sample in Fig. 5(a) exhibits a steeper slope in linewidth vs frequency, corresponding to $\alpha \approx 0.009$. Therefore, the additional damping for this sample is $\Delta\alpha \approx 0.002$. Similar values of $\Delta\alpha$ are obtained for NiFe/Cu/CoFe with epitaxial or polycrystalline CoFe, as well as for NiFe/Cu/Fe with an elemental Fe sink, as shown in Fig. 6 (Cr thickness = 0). This observation is consistent with the (Co)Fe layer acting as a spin absorber, such that a substantial spin current pumped from the NiFe source decays within (Co)Fe. In the following, we use $\Delta\alpha$ as a measure of spin-current absorption by a spin sink – or, equivalently, a measure of spin-current transmission from the spin source to the spin sink. That is, $\Delta\alpha \approx 0.002$ observed for NiFe/Cu/(Co)Fe represents the upper bound for the spin current transmitted through the spacer and absorbed by the sink.

### B. Spin Pumping in Heterostructures with Cu/Cr Spacers

We proceed to examine spin transport in the presence of a thin Cr layer added to the spacer. Figure 5(b-d) presents the frequency dependence of the FMR linewidth for NiFe/Cu/Cr/CoFe, in which Cr and CoFe are epitaxial. Compared to NiFe/Cu/CoFe, we observe a reduced slope in linewidth vs frequency in NiFe/Cu/Cr/CoFe, even with just 0.5 nm of Cr [Fig. 5(b)]. At greater Cr thicknesses [Fig. 5(c,d)], the slope approaches that of the NiFe/Cu sample without a spin sink. Adding a thin Cr layer to the spacer suppresses spin pumping.

Figure 6(a) summarizes the dependence of the spin-pumping damping $\Delta\alpha$ on the epitaxial Cr insertion layer thickness. We observe an approximately tenfold decrease in $\Delta\alpha$ with $\gtrsim$1-nm-thick epitaxial Cr. That is, there is a sharp drop in spin pumping – mostly independent of the Cr thickness – in this sample series with the Cu(001)/Cr(001) spacer [Fig. 6(a)]. This sharp suppression of $\Delta\alpha$ is observed for heterostructures with $Co_{25}Fe_{75}$ alloy and elemental Fe spin sinks. Similar suppression of $\Delta\alpha$ is also obtained with the field applied along the easy and hard axes of epitaxial (Co)Fe [empty and filled symbols, respectively, in Fig. 6(a)]. Thus, we observe no clear anisotropy in the suppression of spin pumping.

We are unable to claim *complete* suppression of spin pumping ($\Delta\alpha \equiv 0$) with Cr insertion. This is due to the sample-to-sample variation in the baseline damping $\alpha_0$, which yields an uncertainty in $\Delta\alpha$ of up to $\approx 2 \times 10^{-4}$ (captured by the error bars in Fig. 6). Nevertheless, we emphasize that the results in Fig. 6(a) demonstrate an order-of-magnitude reduction in spin pumping with Cr added to the Cu spacer.

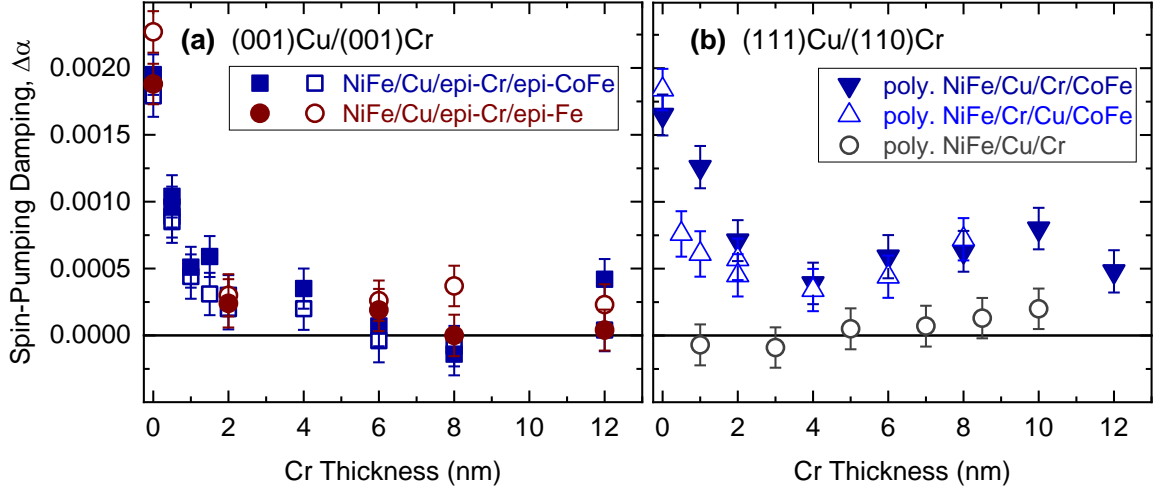

FIG 6. Evolution of the spin-pumping damping Δα with the thickness of the Cr insertion layer in (a) NiFe/Cu/Cr/(Co)Fe heterostructures based on epitaxial Cr and (Co)Fe, with a Cu(001)/Cr(001) spacer, and (b) all-polycrystalline NiFe/Cu/Cr/CoFe heterostructures, with a Cu(111)/Cr(110) spacer. In (a), the filled symbols indicate results obtained with the field applied along the easy axis of the epitaxial (Co)Fe spin sink ($H \parallel$ MAO[110] or (Co)Fe[100]); the empty symbols indicate results obtained with the field applied along the hard axis of the epitaxial (Co)Fe spin sink ($H \parallel$ MAO[100] or (Co)Fe[110]). Note that Δα is a measure of spin current lost from the NiFe spin source (i.e., spin current absorbed in Cu/Cr or CoFe). Δα ≈ 0 for NiFe/Cu/Cr without a CoFe sink, shown in (b), indicates the absence of significant spin absorption in Cu/Cr. The error bars are dominated by the uncertainty ($1.5 \times 10^{-4}$) in the baseline damping $\alpha_0$ that is propagated to Δα [Eq. 1].

A few remarks are in order about the suppressed spin pumping. First, the reduction of Δα to nearly ≈0 indicates that most of the pumped spin current is not absorbed by the (Co)Fe sink. It follows that most spin current is *not transmitted* through the Cu/Cr spacer. Second, any sizable absorption of the spin current (e.g., decoherence via incoherent spin-flip scattering) in the Cu/Cr spacer would result in sizable Δα. The suppression of Δα indicates that most of the spin current is *not absorbed* in the Cu/Cr spacer either.

We investigate whether the suppression of spin pumping is unique to the NiFe/Cu/epi-Cr/epi-(Co)Fe samples with Cu(001)/Cr(001) spacers [Fig. 6(a)]. In Fig. 6(b), we observe that all-polycrystalline NiFe/Cu/Cr/CoFe with a Cu(111)/Cr(110) spacer also exhibits a decline in Δα with Cr insertion. Evidently, spin pumping is reduced in both sample series with different crystallographic orientations.

Yet, the decrease of Δα for the polycrystalline series with the Cu(111)/Cr(110) spacer exhibits a more gradual thickness dependence [Fig. 6(b)], in contrast to the sharp drop for the epitaxial series with the Cu(001)/Cr(001) spacer [Fig. 6(a)]. At large Cr insertion thicknesses, the NiFe/Cu/Cr/CoFe series in Fig. 6(b) {Cu(111)/Cr(110) spacer} retains a systematically higher Δα of ≈ $5 \times 10^{-4}$, compared to the series in Fig. 6(a) {Cu(001)/Cr(001) spacer}. Spin absorption in Cu(111)/Cr(110) is negligible because Δα remains close to zero in NiFe/Cu/Cr samples without a CoFe sink [Fig. 6(b)]. Therefore, the residual Δα ≈ $5 \times 10^{-4}$ in all-polycrystalline NiFe/Cu/Cr/CoFe is attributed to partial spin pumping into the CoFe sink. Overall, we deduce that the polycrystalline Cu(111)/Cr(110) spacer is partially transparent to the spin current, in contrast to the epitaxial Cu(001)/Cr(001) spacer that more strongly suppresses spin pumping. Even with the partially spin-transparent Cu(111)/Cr(110) spacer, we stress that the reduction in spin pumping is still large – i.e., a factor of ≈4 [Fig. 6(b)].

Our above findings reveal that Cu/Cr spacers suppress spin pumping in various NiFe/Cu/Cr/(Co)Fe heterostructures. We have also tested spin pumping in heterostructures with the Cr and Cu spacer layers reversed – i.e., all-polycrystalline NiFe/Cr/Cu/CoFe where the pumped spin current enters Cr first. As shown in Fig. 6(b), the reversed Cr/Cu spacer yields results similar to the Cu/Cr spacer. Hence, the suppressed spin pumping emerges irrespective of whether the spin current enters Cu first or Cr first, in contrast to nonreciprocal spin transport reported for some heterostructures [50].

**C. Origin of the Suppressed Spin Pumping: Bulk vs Interface**

We now wish to address whether the suppression of spin pumping originates from the *bulk* of the Cr insertion layer or the *interface* of Cu/Cr. To this end, we examine spin pumping in NiFe/Cr/CoFe samples with Cu omitted from the spacer [Fig. 7(a)]. In this NiFe/Cr/CoFe series, the Cr thickness is ≥4 nm to minimize interlayer exchange coupling between the NiFe spin source and the CoFe spin sink. We are also limited to all-*polycrystalline* NiFe/Cr/CoFe samples here. As noted in Sec. II-A, NiFe grown directly on top of *epitaxial* Cr exhibits pronounced two-magnon scattering that complicates the interpretation of spin pumping.

As seen in Fig. 7(a), the all-polycrystalline NiFe/Cr/CoFe series exhibits sizable spin-pumping damping of $\Delta\alpha \approx 0.0015$. The NiFe/Cr samples without a CoFe sink [Fig. 7(a)] also exhibit a non-negligible $\Delta\alpha$, suggesting that polycrystalline Cr interfaced directly with the NiFe source may absorb a detectable fraction of the spin current. Additionally, there appears to be a slight increase in $\Delta\alpha$ with Cr thickness in Figs. 6(b) and 7(a), possibly due to the onset of spin absorption in Cr as its thickness approaches the spin diffusion length of ≳10 nm [21]. Nevertheless, the systematically greater $\Delta\alpha$ for NiFe/Cr/CoFe compared to NiFe/Cr (by a factor of ≳ 2) indicates that a large fraction (≳ 50%) of the spin current is transmitted across the Cr spacer (and absorbed in the CoFe sink).

As an additional check of spin transport through the single-layer Cr spacer, we have performed an x-ray synchrotron-based spin pumping experiment [9,31,51,52] on NiFe/Cr/CoFe at Beamline 4.0.2 of the Advanced Light Source, Lawrence Berkeley National Laboratory. The sample for this experiment was grown on a MgO substrate to allow for luminescence yield detection of X-ray magnetic circular dichroism (XMCD). The details of this experimental setup are found in Refs. [52,53]. In brief, XMCD is used to detect the magnetization dynamics (i.e., magnetization component transverse to the precessional axis) of a specific element. For instance, we acquire the in-plane field dependence of the precessional amplitude and phase for Ni in the NiFe source, driven resonantly by a 3-GHz microwave. As shown in Fig. 7(b), a peak in the amplitude and a 180-degree shift in the phase are observed for Ni, consistent with the FMR of the NiFe source. In addition, we detect the Co magnetization dynamics in the CoFe sink near the resonance field of NiFe, indicating dynamic coupling between the NiFe source and the CoFe sink [9]. The data for the Co dynamics are adequately fitted with a model based on coupled Landau-Lifshitz-Gilbert equations [9,31,51,52], as shown in Fig. 7(b). This model accounts for the off-resonant microwave field torque (appearing as the non-zero offset in the amplitude in Fig. 7(b)), interlayer dipolar field torque (green dashed curves in Fig. 7(b)), and spin torque driven by the spin current pumped into CoFe (red solid curves in Fig. 7(b)). Of particular note here is the spin torque, signifying sizable spin transmission from the NiFe source to the CoFe sink [9,31,51,52]. Hence, this synchrotron-based experiment corroborates that the single-layer Cr spacer is indeed transparent to the spin current.

Our complementary results in Fig. 7 indicate spin pumping through single-layer Cr spacers. At the same time, our findings in Sec. II-B demonstrate that spin pumping is suppressed in heterostructures with bilayer

Cu/Cr spacers. We therefore identify the Cu/Cr interface, rather than the bulk of Cr, as the origin of the suppressed spin pumping.

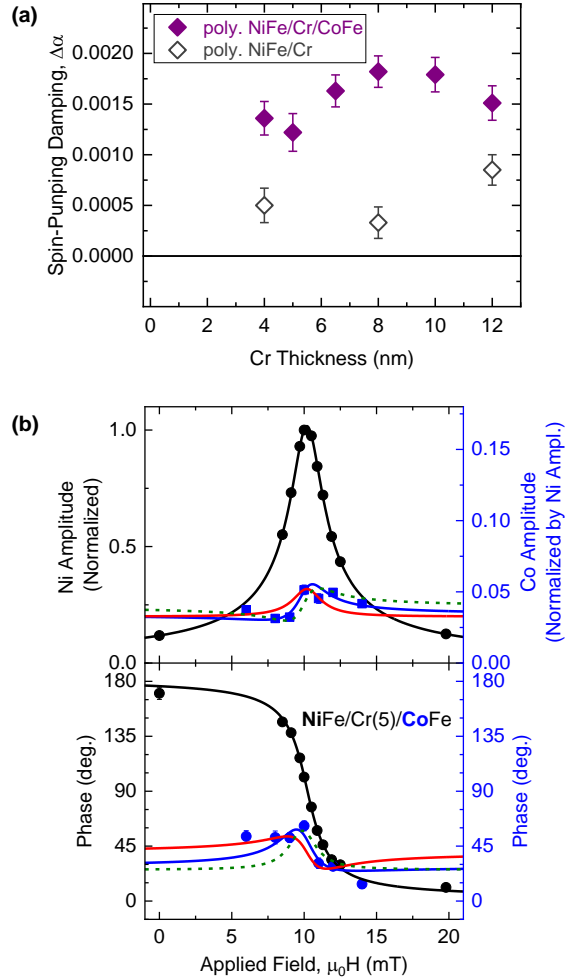

FIG 7. (a) Evolution of the spin-pumping damping $\Delta\alpha$ with the thickness of the single-layer Cr spacer in all-polycrystalline NiFe/Cr/CoFe (filled symbols), as well as NiFe/Cr without a CoFe sink (empty symbols). The error bars are dominated by the uncertainty ($1.5 \times 10^{-4}$) in the baseline damping $\alpha_0$ that is propagated to $\Delta\alpha$ [Eq. 1]. (b) Precessional amplitude and phase of the Ni and Co magnetizations in NiFe/Cr/CoFe (Cr thickness 5 nm), measured with XMCD. Accompanying the Co results (blue data points), the solid blue fit curves represent the total torque acting on the Co magnetization, whereas the solid red (dashed green) fit curves represent the contribution from the spin torque (interlayer dipolar field torque).

**D. Interpretation and Fundamental Mechanism of the Suppressed Spin Pumping**

It is quite surprising that combining Cu and Cr in the spacer suppresses spin transmission, particularly given that thin Cu and Cr by themselves are transparent to spin currents. Both Cu and Cr are electrically conductive 3$d$ transition metals with weak spin-orbit coupling, which would be expected to permit efficient spin transmission. These points are consistent with our findings of spin pumping through a thin single-layer Cu or Cr spacer with a thickness well below the spin diffusion length. Yet, interfacing Cu with just a few monolayers of Cr drastically reduces spin pumping through the spacer [Fig. 6].

Explaining the suppression of spin pumping is complicated because the underlying theoretical mechanism likely extends beyond the Cu/Cr interface – even though, experimentally, this particular interface appears to cause the suppression. Here, we use a simple two-channel model in Fig. 8 to illustrate the deficiency of the theory that focuses solely on the Cu/Cr interface. In this model, $\Delta\mu_\sigma = \mu_{Cu,\sigma} - \mu_{Cr,\sigma}$ denotes the nonequilibrium chemical potential difference across the interface for each spin direction ($\sigma = \uparrow$ or $\downarrow$). $R$ gives the interfacial resistance for each spin channel that represents carrier flow for each spin. Due to the lack of ferromagnetism at the interface, both spin channels must have an identical interface resistance $R$, regardless of the presence of spin-orbit coupling or antiferromagnetism in Cr. Since a pure spin current is represented by the spin channels having equal and opposite currents (i.e., $\Delta\mu_\uparrow = -\Delta\mu_\downarrow$), pure-spin-current transport decreases only when the interfacial resistance $R$ increases equally for both spin channels. That is, large spin-pumping suppression in the Cu/Cr system can be replicated only under the implausible condition that the metallic Cu/Cr interface blocks electronic charge transport. Thus, the theoretical model of the Cu/Cr interface alone cannot capture the observed suppression of spin pumping.

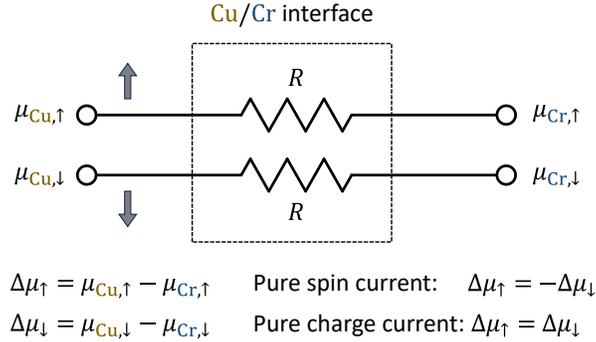

FIG 8. Schematic of the two-channel model of the Cu/Cr interface, consisting of spin-up and spin-down channels. Both the pure spin and charge currents are driven by a difference in the chemical potentials ($\Delta\mu_\uparrow$ and $\Delta\mu_\downarrow$) across the interface. At this interface of two non-ferromagnetic metals, the interfacial resistance $R$ must be equal for both spin channels. To suppress spin pumping through this interface, the charge resistance at the interface must diverge – which would be an unlikely scenario for the metallic Cu/Cr interface. Therefore, this simple two-channel model of the Cu/Cr interface is unable to provide a plausible explanation for the suppression of spin pumping.

A possible explanation for the spin-pumping suppression is a large reduction in the spin-mixing conductance [54], e.g., that encompasses the NiFe/Cu/Cr system. Conventionally, the spin-mixing conductance $G_{\uparrow\downarrow}$ is a parameter describing a ferromagnet/non-ferromagnet (FM/NM) interface [54]; $G_{\uparrow\downarrow}$ relates the transverse spin chemical potential $\vec{\mu}_t$ to the transversely-polarized spin current $\vec{j}_t$ on the NM side of the interface ($\vec{j}_t \propto G_{\uparrow\downarrow}\vec{\mu}_t$), where "transverse" is defined relative to the magnetization in the FM. A smaller spin-mixing conductance would result in a smaller spin current (spin pumping) in the heterostructure. In the absence of spin-orbit coupling, the spin-mixing conductance depends solely on the reflection amplitudes of electrons scattering off the FM/NM interface. However, if another NM' layer is inserted between the original FM and NM layer to constitute a FM/NM'/NM system (e.g., NiFe/Cu/Cr), the effective spin-mixing conductance could be modified, potentially due to coherent backscattering within the inserted NM' layer. The NiFe/Cu/Cr system may exhibit a much smaller effective spin-mixing conductance – compared to the NiFe/Cu or NiFe/Cr system – that greatly reduces spin pumping in the heterostructure. While quantitative calculations of the spin-mixing conductance are beyond the scope of this present work, the large modification of spin pumping in FM/NM'/NM systems warrants further theoretical studies.

Prior experimental studies [55,56] have reported modifications of the spin-mixing conductance by inserting a thin additional NM' layer in a FM/NM bilayer. However, the modifications in these studies are limited to a factor of $\approx 2$. With the spin-mixing conductance proportional to spin-pumping damping $\Delta\alpha$, the modifications seen in our present study are far greater – i.e., an order of magnitude reduction in the spin-mixing conductance. Such a giant reduction is reminiscent of suppressed spin pumping by inserting thin nonmagnetic insulators [57,58]. We rule out the possibility of oxidized Cr impeding spin transport, considering the low background pressure in the deposition chamber during and after the growth of Cr and the XRD results with no evidence for oxidized Cr [see Sec. II]. Moreover, sizable spin transport takes place through single-layer Cr spacers [Sec. III-C], so Cr alone cannot account for the suppression of spin pumping. A nontrivial, previously unexplored mechanism is likely responsible for the suppressed spin pumping at metal interfaces.

### E. Spin Pumping in Heterostructures with Other Bilayer Spacers

The initial motivation of our work was to examine the influence of elemental antiferromagnetic Cr on interlayer spin transport. It is sensible to inquire whether the antiferromagnetism of Cr is responsible for suppressing spin pumping at the Cu/Cr interface. To address this question, we have investigated spin pumping in heterostructures with alternative Cu/$X$ spacers, i.e., where $X$ is a nonmagnetic transition metal, here V or Ag.

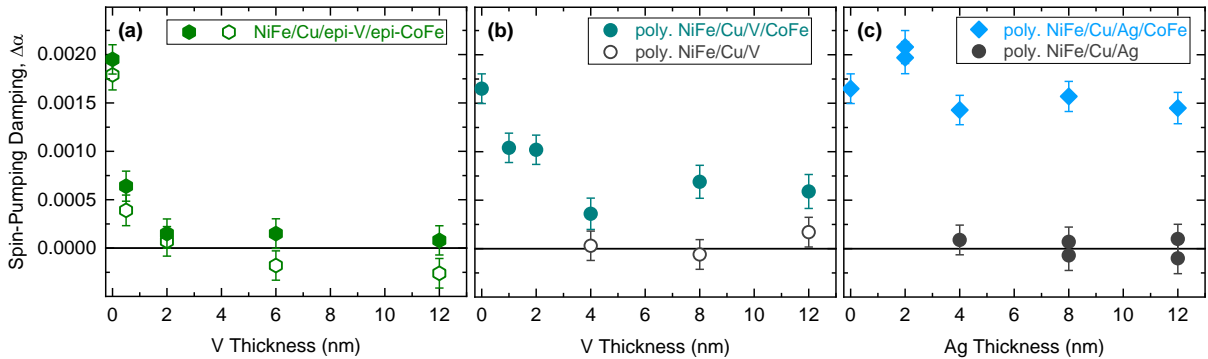

FIG 9. (a,b) Evolution of the spin-pumping damping $\Delta\alpha$ with the thickness of the V insertion layer in (a) NiFe/Cu/V/CoFe heterostructures based on epitaxial V and CoFe and (b) all-polycrystalline NiFe/Cu/V/CoFe heterostructures. In (a), the filled symbols indicate results obtained with the field applied along the easy axis of the epitaxial CoFe spin sink ($H \parallel$ MAO[110] or CoFe[100]); the empty symbols indicate results obtained with the field applied along the hard axis of the epitaxial CoFe spin sink ($H \parallel$ MAO[100] or CoFe[110]). (c) Evolution of the spin-pumping damping $\Delta\alpha$ with the thickness of the Ag insertion layer in all-polycrystalline NiFe/Cu/V/CoFe heterostructures. The error bars are dominated by the uncertainty ($1.5 \times 10^{-4}$) in the baseline damping $\alpha_0$ that is propagated to $\Delta\alpha$ [Eq. 1].

We first present spin-pumping results for heterostructures with Cu/V spacers in place of Cu/Cr. The comparison between Cu/V and Cu/Cr is interesting because V and Cr are structurally similar. The atomic number $Z = 23$ of V neighbors $Z = 24$ of Cr, and both V and Cr are BCC crystals with similar bulk lattice parameters (0.303 nm and 0.291 nm, respectively). In effect, Cu/V is a non-antiferromagnetic analogue of Cu/Cr.

Figure 9(a,b) summarizes the FMR spin-pumping results for two series of heterostructures: (1) those incorporating *epitaxial* V, grown on top of epitaxial (Co)Fe on (001)-oriented MAO [Fig. 9(a)] and (2) those incorporating *polycrystalline* V, grown on top of other polycrystalline film layers on Si-SiO$_2$ [Fig. 9(b)]. As seen in Fig. 9(a), the insertion of epitaxial V in the spacer sharply decreases the spin-pumping damping $\Delta\alpha$ to $\approx 0$. This observation resembles the sharp decline in $\Delta\alpha$ with inserting epitaxial Cr in Fig. 6(a). The all-polycrystalline samples in Fig. 9(b) also show a decrease in $\Delta\alpha$ with V insertion, down to $\Delta\alpha \approx 5 \times 10^{-4}$ – again, akin to the results with Cr insertion [Fig 6(b)]. We also see negligible spin-pumping damping in NiFe/Cu/V (without a CoFe sink), indicating that Cu/V does not significantly absorb the pumped spin current. Taken together, the observed trends here for the Cu/V-based heterostructures [Fig. 9(a,b)] are remarkably similar to those for the Cu/Cr-based heterostructures [Fig. 6]. Our results indicate that Cr and V, when interfaced with Cu to comprise a bilayer spacer, have essentially the same effect on spin transport. Antiferromagnetic Cr is not required for the suppression of spin pumping.

We have thus identified two bilayer spacers (Cu/Cr and Cu/V) that suppress spin pumping. It is then instructive to determine whether *any* bilayer spacer of Cu/*X* can suppress spin pumping. To this end, we have investigated heterostructures incorporating bilayer Cu/Ag spacers. As shown in Fig. 9(c), the spin-pumping damping $\Delta\alpha$ is *not* suppressed with the addition of Ag to the spacer. The control series of NiFe/Cu/Ag without a CoFe shows $\Delta\alpha \approx 0$, which corroborates that the large $\Delta\alpha$ in NiFe/Cu/Ag/CoFe originates from spin pumping into CoFe, i.e., through Cu/Ag. That is, the bilayer Cu/Ag spacer is just as transparent to the spin current as the single-layer Cu spacer. We conclude that while the suppression of spin pumping is not unique to heterostructures with Cu/Cr spacers, it is not universal to all heterostructures with bilayer Cu/*X* spacers.

A crystal-structure mismatch between the two metals in the bilayer spacer may be crucial for suppressing spin pumping. Namely, FCC Cu interfaced with BCC Cr or V suppresses spin pumping, whereas FCC Cu interfaced with FCC Ag does not. It is possible that the mismatch in crystal structure – hence electronic band structures – affects the effective spin-mixing conductance of the heterostructure. The difference in the Fermi energy or carrier effective mass between the two metals could impede the propagation of Bloch wave packets, which fundamentally govern electronic spin transport. Nevertheless, since our present study examines only limited combinations of metals, the possible role of crystal and electronic structure mismatch remains speculative. How a thin metallic insertion layer decreases spin pumping – e.g., by an order of magnitude – remains an open question that requires further experimental and theoretical work.

### F. Temperature Dependence of Spin Pumping

All the above results [Secs. III-A through III-E] are obtained from experiments at room temperature. The Cr layers studied here may exhibit some antiferromagnetic order at room temperature, considering its bulk ordering temperature of 311 K. Even for the small thicknesses of Cr, the ordering temperature could remain close to the bulk limit due to the proximity to ferromagnetic (Co)Fe [59]. At lower temperatures, the antiferromagnetic order should become stronger and, particularly for crystalline Cr, may exhibit rich physics associated with spin-density waves [12,13]. Therefore, to examine the possible influence of stronger antiferromagnetic order on spin transport, we have performed variable-temperature experiments.

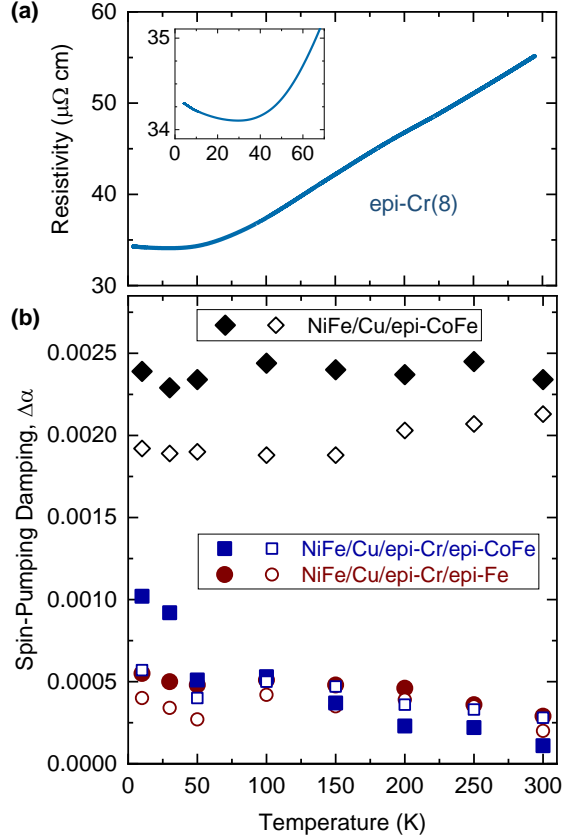

FIG 10. (a) Temperature dependence of the electrical resistivity of an 8-nm-thick epitaxial Cr film. Inset: uptick of the resistivity with decreasing temperature below 30 K. (b) Temperature dependence of $\Delta\alpha$ for NiFe/Cu/epi-Cr/epi-(Co)Fe heterostructures, with Cr thickness 8 nm. The filled symbols indicate results obtained with the field applied along the easy axis of the epitaxial (Co)Fe spin sink ($H \parallel$ MAO[110] or (Co)Fe[100]); the empty symbols indicate results obtained with the field applied along the hard axis of the epitaxial (Co)Fe spin sink ($H \parallel$ MAO[100] or (Co)Fe[110]).

While determining the antiferromagnetic configurations is beyond the scope of our present work, we are able to gain partial insights into the antiferromagnetic order in Cr films through the temperature dependence of electrical resistivity. Figure 10(a) presents resistivity vs temperature for an 8-nm-thick epitaxial Cr film grown directly on MAO. The monotonic decrease in resistivity with decreasing temperature, down to ≈30 K, is consistent with the metallic nature of Cr. However, the resistivity shows a slight uptick with further reduction in temperature below ≈30 K. This uptick can be due to several mechanisms, including: (1) Anderson (strong) localization due to lattice disorder, described by the variable range hopping model [60–63]; (2) Efros-Shklovskii localization, where electron-electron interactions open a gap at the Fermi energy [60,62]; (3) the spin Kondo effect [64]; (4) weak-localization with a carrier dephasing time limited by electron-electron quasi-elastic Nyquist scattering (Altshuler-Aronov effect) [60–63]; (5) an exchange/Hartree correction to the resistivity due to effects of electron-electron interactions on the density of states [60,63]; 6) resonant impurity scattering in metallic antiferromagnets, which has been reported in antiferromagnetic Cr films [65]. Of these mechanisms, (6) appears the most likely. Mechanisms (1) and

(2) result in an exponential dependence on temperature at low temperatures in contrast to the weak uptick in resistivity observed in Fig. 10(a). Mechanism (3) is unlikely since the spin Kondo effect occurs from scattering of carriers by magnetic impurities typically in metals with dilutely dispersed magnetic impurities. Cr in contrast has non-zero magnetic moment at each lattice atom, and a spin Kondo effect is not likely to manifest in such concentrated magnetic system; Ref. [65] arrives at the same conclusion. Mechanisms (4) and (5) are viable alternatives to the effects of resonant impurity scattering in antiferromagnets (6). The data does not allow a fully unambiguous distinction, since various models can be fitted to reproduce the data fairly well. Yet, the strong similarity between Fig. 10(a) and the data in [65] (resonant impurity scattering in antiferromagnetic Cr), in shape and magnitude of the uptick in resistivity and in the temperature range where it manifests, makes resonant impurity scattering the most likely explanation. Thus, we deduce that Fig. 10(a) supports the evidence that the Cr thin film is indeed antiferromagnetic at such low temperatures.

We have also conducted variable-temperature FMR spin-pumping measurements [Fig. 10(b)], employing a spectrometer equipped with a cryostat, for heterostructures grown on epitaxial (Co)Fe. We use the temperature dependence of the intrinsic damping parameter $\alpha_0$ of NiFe (measured from a control NiFe/Cu sample without CoFe or Cr) as the baseline to quantify the temperature dependence of damping enhancement $\Delta\alpha$. The NiFe/Cu/CoFe sample shows a large, nearly constant $\Delta\alpha$ of $\approx 0.002$ across the entire temperature range. For this sample, the values of $\Delta\alpha$ are systematically higher by $\approx 20\%$ for measurements with the field applied along the easy axis of CoFe [filled symbols in Fig. 10(b)]. We speculate that this apparent anisotropy is due to small two-magnon scattering or anisotropic spin pumping [31].

For the NiFe/Cu/Cr/(Co)Fe samples, $\Delta\alpha$ remains small, i.e., $< 5 \times 10^{-4}$, across the entire temperature range. There appears to be a slight increase of $\Delta\alpha$ with decreasing temperature, although it is difficult to discern a clear trend from the scatter in the data. The antiferromagnetic order of Cr, which becomes stronger at lower temperatures, evidently has little impact on spin pumping. Yet, at the low-temperature limit, we observe an abrupt increase in $\Delta\alpha$ up to $\approx 0.001$ for the NiFe/Cu/Cr/CoFe sample, measured with the field along the easy-axis of CoFe. While the origin of this abrupt increase for that particular sample (and the particular measurement geometry) is unknown, no such increase is seen for the similar NiFe/Cu/Cr/Fe sample. Therefore, we conclude that the antiferromagnetic order of Cr in of itself does not significantly influence spin transport in these heterostructures.

## IV. CONCLUSIONS

By employing FMR spin pumping, we have studied pure-spin-current transport in metallic heterostructures that incorporate the elemental antiferromagnet of Cr. We have primarily focused on heterostructures of the form NiFe/Cu/Cr/(Co)Fe, where the Cu/Cr spacer separates the NiFe spin source and the (Co)Fe spin sink. We find that the Cu/Cr spacer greatly reduces spin pumping – i.e., neither transmitting nor absorbing a significant amount of spin current. This suppression of spin pumping is rather surprising, considering that a thin layer of Cu or Cr alone permits significant spin transmission. A particularly large suppression (i.e., by an order of magnitude) emerges at the interface of Cu(001)/Cr(001), although the interface of Cu(111)/Cr(110) also yields a sizable reduction (by a factor of $\approx 4$). Moreover, we observe similar suppression of spin pumping with Cu/V spacers, where V is a nonmagnetic analogue of Cr, demonstrating that the antiferromagnetism of Cr is not responsible for suppressing spin pumping. While spin pumping is suppressed with FCC/BCC spacers of Cu/Cr and Cu/V, no suppression arises with FCC/FCC spacers of Cu/Ag. The mismatch of crystal structure – hence electronic band structure – at the interface of non-ferromagnetic metals may play a critical role, e.g., in the effective spin-mixing conductance. Finally, the antiferromagnetism of Cr does not appear to impact spin transport strongly in NiFe/Cu/Cr/(Co)Fe over a wide temperature range of 10-300 K. Our work may stimulate a new outlook on spin transport in metallic systems, including interfaces that are electrically conductive and yet spin insulating.

**ACKNOWLEDGEMENTS**

Y.L., D.A.S., J.J.H., and S.E. were supported by the National Science Foundation (NSF) under Grant No. DMR-2003914. B.N. was supported by NSF MEMONET Grant No. 1939999. C.K. and P.S acknowledge support from the US Department of Energy, Office of Science, Office of Basic Energy Sciences, the Microelectronics Co-Design Research Program, under contract no. DE-AC02-05-CH11231 (Codesign of Ultra-Low-Voltage Beyond CMOS Microelectronics). V.P.A., T.M., and I.-J.P. acknowledge support by the NSF under Grant No. DMR-2105219. This work was made possible by the use of Virginia Tech's Materials Characterization Facility, which is supported by the Institute for Critical Technology and Applied Science, the Macromolecules Innovation Institute, and the Office of the Vice President for Research and Innovation. This research used resources of the Advanced Light Source, a U.S. DOE Office of Science User Facility under Contract No. DE-AC02-05CH11231. We thank Eric Montoya and Kyungwha Park for helpful discussions.

**DATA AVAILABILITY**

The data that support the findings of this study are available from the corresponding authors upon reasonable request.